\documentclass[12pt,showpacs,amsfonts]{revtex4-1}
\usepackage{amsmath}
\usepackage{latexsym}
\usepackage{float}
\usepackage{amssymb}
\usepackage{graphicx}
\usepackage{textcomp}
\usepackage{hyperref}
 1
\def\ba{\begin{eqnarray}}
\def\ea{\end{eqnarray}}
\def\be{\begin{equation}}
\def\ee{\end{equation}}
\def\bm{\begin{math}}
\def\me{\end{math}}

\newcommand{\dummy}

\begin{document}
\title{Kinetics of Fluid Phase Separation}
\author{{Subir K. Das}*, Sutapa Roy and Jiarul Midya}
\affiliation{Theoretical Sciences Unit, Jawaharlal Nehru Centre for Advanced Scientific Research,
 Jakkur P.O., Bangalore 560064, India}
\date{\today}

\begin{abstract}
We review understanding of kinetics of fluid phase separation in various space dimensions. Morphological differences, 
percolating or disconnected, based on overall composition in a binary liquid or density in a vapor-liquid system, have been 
pointed out. Depending upon the morphology, various possible mechanisms and corresponding theoretical predictions for 
domain growth are discussed. On computational front, useful models and simulation methodologies have been presented. 
Theoretically predicted growth laws have been tested via molecular dynamics simulations of vapor-liquid transitions. In 
case of disconnected structure, the mechanism has been confirmed directly. This is a brief review on the topic for a special 
issue on coarsening dynamics, expected to appear in Comptes Rendus Physique. 
\end{abstract} 
\maketitle
\section{Introduction}
~Topics related to phase transitions received significant attention over many decades 
\cite{MFisher1,HEStanley,REvans,KBinder1,KBinder2,DKashchiev,RALJones1,AOnuki,AJBray,SPuri1,SKDas1}. 
The objective of this review is to discuss developments in the understanding of kinetics of phase transitions
\cite{MFisher1,HEStanley,REvans,KBinder1,KBinder2,DKashchiev,RALJones1,AOnuki,AJBray,SPuri1,SKDas1,IMLifshitz,DAHuse1,
JFMarko,DWHeermann,JVinals,SMajumder1,SMajumder2,TBlanchard,KBinder3,KBinder4,EDSiggia,HFurukawa1,HFurukawa2,MSMiguel,
HTanaka1,HTanaka2,HTanaka3,FPerrot,JPDelville,JHobley,DBeysens,STanaka,VMKendon,SPuri2,CDatt,MLaradji,AKThakre,SAhmad1,
SAhmad2,SAhmad3,SKDas2,HKabrede,SMajumder3,SRoy1,SRoy2,SRoy3}, with emphasis on phase separating fluid systems 
\cite{KBinder3,KBinder4,EDSiggia,HFurukawa1,HFurukawa2,MSMiguel,HTanaka1,HTanaka2,HTanaka3,FPerrot,JPDelville,JHobley,
DBeysens,STanaka,VMKendon,SPuri2,CDatt,MLaradji,AKThakre,SAhmad1,SAhmad2,SAhmad3,SKDas2,HKabrede,SMajumder3,SRoy1,SRoy2,
SRoy3,SWKoch,JMidya}. Evolution of a system from one equilibrium phase to another is a nonequilibrium phenomenon 
and happens via nucleation \cite{MFisher1,HEStanley,REvans,KBinder1,KBinder2,DKashchiev} and growth 
\cite{KBinder2,RALJones1,AOnuki,AJBray,SPuri1,SKDas1} of domains. E.g., when a homogeneous binary mixture ($A+B$), 
liquid or solid, is quenched inside the miscibility gap, it becomes unstable to fluctuations and moves towards the 
coexisting phase-separated state via formation and growth of domains rich in $A$ and $B$ particles. For a vapor-liquid 
transition, the approach to the new equilibrium occurs via formation of particle rich and particle poor regions.
\par
~Typically, the growth of domains is a scaling phenomena \cite{AJBray}, i.e., there is self-similarity of patterns at 
two different times, despite a change in length scale $\ell(t)$, the average size of domains at time $t$. This is 
captured in the scaling properties \cite{AJBray}
\begin{eqnarray}\label{scl_corr}
C(r,t)\equiv \tilde{C}[r/\ell(t)],
\end{eqnarray}
\begin{eqnarray}\label{scl_stfac}
S(k,t)\equiv \ell(t)^{d} \tilde{S}[k\ell(t)],
\end{eqnarray}
\begin{eqnarray}\label{scl_pd}
P(\ell_{d},t)\equiv \ell(t)^{-1} \tilde{P}[\ell_{d}/\ell(t)],
\end{eqnarray}
of the two-point equal time correlation function ($C$), structure factor ($S$) and domain size distribution function ($P$).
In Eqs. (\ref{scl_corr}-\ref{scl_pd}), $r$ is the distance between two points ($\vec{r_i}$ and 
$\vec{r_j}$), $k$ is the wave vector, $\ell_d$ is the size of a domain and $d$ is the system dimensionality. 
There $\tilde{C}$($x$), $\tilde{S}$($y$) and $\tilde{P}$($z$) are master functions, independent of time.
The above mentioned correlation function is calculated as \cite{AJBray}
\begin{eqnarray}\label{def_corr}
C(r,t)=<\psi(\vec{r_i},t)\psi(\vec{r_j},t)>-<\psi(\vec{r_i},t)><\psi(\vec{r_j},t)>,
\end{eqnarray}
where $\psi$ is the relevant order-parameter for the transition. $S(k,t)$ is the Fourier transform of $C(r,t)$. The scalar 
notations $r$ and $k$ imply spherical isotropy and are applicable for systems without any bias such that the structures are 
isotropic in statistical sense. The angular brackets in Eq.(\ref{def_corr}) are for statistical averaging.
\par
~ Typically, $\ell(t)$ grows with time in power-law fashion as \cite{AJBray}
\begin{eqnarray}\label{leng_scl}
\ell(t) \sim t^{\alpha}.
\end{eqnarray}
The growth exponent $\alpha$ depends upon $d$, number of order-parameter components, morphology, hydrodynamic 
effects and conservation of order-parameter \cite{AJBray}. In out of equilibrium systems, as is clear from 
Eq. (\ref{def_corr}), the order parameter $\psi$ is a function of space and time. In a magnetic system this can be 
identified with the local magnetization, in a binary mixture with the local difference between the concentration of 
the two species, for a vapor-liquid system with the density. Depending upon the type of transition, the total 
order-parameter (the local value integrated over the whole system) may or may not remain same at all times. 
E.g., in a paramagnetic to ferromagnetic transition, where, starting from a zero net magnetization, the system 
gets spontaneously magnetized, the order-parameter is a nonconserved quantity. On the other hand, for all types 
of phase separating systems, assuming that there is no chemical conversion or flow of material between the system 
and  a reservoir, this is a conserved quantity. In this paper we will deal with conserved scalar order parameter.
\par
~Major fraction of the literature in kinetics of phase transitions is related to the behavior of $C(r,t)$ and 
understanding of the exponent $\alpha$. However, there exist other interesting aspects, e.g., aging \cite{DSFisher1} and 
persistence \cite{SNMajumdar}. In aging phenomena, typically one studies the two-time quantities, e.g., 
the order-parameter autocorrelation function \cite{DSFisher1}
\begin{eqnarray}\label{def_autocorr}
C_{\rm{age}}(t,t_w)=<\psi(\vec{r_i},t)\psi(\vec{r_i},t_w)>-<\psi(\vec{r_i},t)><\psi(\vec{r_i},t_w)>,
%\nonumber\\*
\end{eqnarray}
where $t_w$ and $t$ ($>t_w$) are respectively the waiting and observation times. Even though the time translation 
invariance is not obeyed by $C_{\rm{age}}(t,t_w)$ in out of equilibrium systems, this quantity is expected to exhibit 
scaling \cite{DSFisher1} with respect to $t/t_w$. An objective in studies related to aging is to obtain scaling 
functions for $C_{\rm{age}}$ or for other two-time quantities of relevance, in different types of phase transitions. 
Persistence, on the other hand, is related to the time dependence of the fraction of unaffected local order parameter. 
\par
~ Each of these aspects are better understood in nonconserved order-parameter case \cite{AJBray}. For conserved 
order-parameter, significant progress has been made only with respect to the growth exponent $\alpha$, 
for solid binary mixtures \cite{IMLifshitz,DAHuse1,JFMarko,DWHeermann,JVinals,SMajumder1,SMajumder2}. 
For the latter, diffusion (via evaporation and condensation) is the only transport mechanism and the growth 
is characterized by a single exponent. In fluids, however, no single exponent describes the entire 
growth process \cite{AJBray}. This is due to the faster transport of material, particularly due to the influence 
of hydrodynamics, at late times. Effects of such fast transport may be manifested in different ways depending 
upon system dimensionality and domain pattern. Thus, kinetics of phase separation in fluids is a richer and 
challenging area. In spite of that, reasonable progress has been made 
\cite{KBinder3,KBinder4,EDSiggia,HFurukawa1,HFurukawa2,MSMiguel,HTanaka3,FPerrot,JPDelville,JHobley,DBeysens,STanaka,
VMKendon,SPuri2,CDatt,MLaradji,AKThakre,SAhmad1,SAhmad2,SAhmad3,SKDas2,HKabrede,SMajumder3,SRoy1,SRoy2,SRoy3,SWKoch,JMidya}. 
The aim of the article is to review some of these works in brief.
\par 
~The rest of the article is organized as follows. In section II we discuss the theoretical progress. Section III is 
devoted to the discussion of models and computational methods. Representative computational results are presented in 
section IV. Finally, section V concludes the paper with a brief summary and outlook.
\section{Theories}
~In a solid binary mixture, the rate of change of domain size is associated with the chemical potential gradient as 
\cite{DAHuse1} $d\ell/dt \sim |\vec{\nabla}\mu|$. Using the dimensionality of $\mu$ as $\gamma/\ell$, $\gamma$ being 
the interfacial tension and assuming that a gradient exists over the length of the domain size, one obtains \cite{DAHuse1}
 \begin{eqnarray}\label{diff_domain}
\frac{d\ell}{dt} \sim \frac{\gamma}{\ell^2}.
\end{eqnarray}
Solution of Eq.(\ref{diff_domain}) provides $\alpha=1/3$, referred to as the Lifshitz-Slyozov (LS) law \cite{IMLifshitz} 
and is understood to be valid for any domain growth occurring via simple diffusive mechanism.
\par
~Early part of fluid phase separation is also dominated by diffusion and thus, is expected to have $\alpha=1/3$. 
At late times, hydrodynamic effects become more important, for percolating domain morphology. As already mentioned, 
while for phase separation in solid mixtures system dimensionality and domain morphology do not play important 
role \cite{SMajumder2}, these become crucial for selection of mechanism in fluids. In some cases, different 
mechanisms may give rise to same value of the growth exponent. Thus, very direct computational or 
experimental probes are necessary to validate a particular one. Below we will discuss some relevant mechanisms briefly.

\begin{figure}[htb]
\centering
\includegraphics*[width=0.45\textwidth]{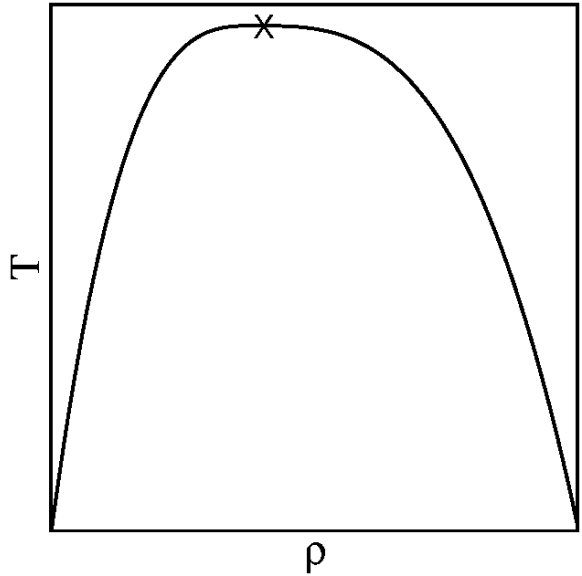}
\caption{\label{fig1} Schematic diagram for vapor-liquid coexistence in $T$ vs $\rho$ plane. The critical point
is marked by a cross. The left and right branches represent respectively the vapor and the liquid densities.
From Molecular simulation.}
\end{figure}

\par
~Typically, phase separating systems exhibit interconnected  or droplet morphology, depending upon the 
overall composition or density of the system. In Fig.\ref{fig1} we show a vapor-liquid coexistence curve 
in temperature ($T$) versus density ($\rho$) plane. There the left branch corresponds to the vapor-phase 
density and the right one to the liquid-phase density. For a binary mixture phase separation, 
the appropriate abscissa variable is the concentration of one of the species. 
For density or composition very close to the coexistence curve (vapor branch in vapor-liquid transition), 
droplet like pattern is visible, from the very beginning. In this under-saturation case, 
the system requires fluctuations involving large length scales to nucleate stable droplets that can grow. 
Such events, depending upon the proximity to the coexistence curve, may be rare and the phase 
separation can be delayed. On the other hand, far from the coexistence curve, the system falls 
unstable to infinitesimal fluctuations and the corresponding phase separation process is referred 
to as the spinodal decomposition. Typically, in this case, the domain morphology is percolating. 
Here we caution the reader that at late time droplet structures can emerge from percolating morphology 
as well \cite{CDatt}, if the system is away from the critical composition or density. In fluids this 
is expected to happen faster.
\par
~Let us first consider the case of percolating domain morphology. Because of experimental realizability, 
we will first take up the $d=3$ case. The hydrodynamic regime is divided into two subregimes \cite{AJBray}, 
viz., viscous and inertial. These we discuss below, following Siggia \cite{EDSiggia} and Furukawa 
\cite{HFurukawa1,HFurukawa2}, using the Navier-Stokes (NS) equation
\begin{eqnarray}\label{stokes_eq}
\rho \frac{D}{Dt}\vec{v}-\nu \nabla^2 \vec{v}=-\vec{\nabla}p.
\end{eqnarray}
In Eq.(\ref{stokes_eq}), $\rho$ is the mass density, $\vec{v}$ is the fluid velocity, $\nu$ is the kinematic viscosity, 
$p$ is the pressure and $D/Dt= d/dt+(\vec{v}.\vec{\nabla})$. In $D\vec{v}/Dt$, both the terms stand for acceleration, 
as can be checked dimensionally. The second one is related to convection where the acceleration is generated 
by space dependent velocity. Eq.(\ref{stokes_eq}) is essentially related to balancing of force 
coming from pressure with that of frictions. On the left hand side, the first term can be treated as inertial friction and 
the second one is related to dissipative friction originating from viscous drag. Via dimensional substitutions 
$Dv/Dt\equiv v/t$, $\nabla^{2}v\equiv -v/\ell^2$ and considering that the pressure gradient is obtainable from 
interfacial tension $\gamma$, one writes, after multiplying both sides by $vt$,
\begin{eqnarray}\label{stokes_eq2}
\rho v^2 + \frac{\rho \nu}{\ell} v=\frac{\gamma}{\ell}.
\end{eqnarray}
The second term on the left hand side of Eq.(\ref{stokes_eq2}) is related to the inverse of time dependent Reynolds number 
and is important at earlier part of the hydrodynamic regime, providing viscous growth. At later time, the first term 
dominates, giving rise to the inertial hydrodynamic growth. Thus, in the first part of hydrodynamic growth it is equivalent 
to equating the interfacial free energy density with the viscous stress and in the second part, the same with the kinetic 
energy density \cite{EDSiggia}. With the understanding that there exist unique time and length scales in the problem, one 
identifies $v$ with $d\ell/dt$. Then, in the viscous regime one obtains
\begin{eqnarray}\label{domain_vel}
\frac{d\ell}{dt} = \frac{\gamma}{\rho\nu},
\end{eqnarray}
and in the inertial regime
\begin{eqnarray}\label{domain_vel2}
\frac{d\ell}{dt}=\sqrt{\frac{\gamma}{\rho\ell}}.
\end{eqnarray}
Solutions of Eqs. (\ref{domain_vel}) and (\ref{domain_vel2}) provide $\alpha=1$ and $2/3$, respectively.
\par
~Recall that our discussion is related to interconnected domain morphology, having tube-like structure in $d=3$. The 
fast advective  transport of material through these tubes happens due to pressure gradient calculable from interfacial 
tension and domain size. In $d=2$, analogue of a tube is strip. San Miguel et al. \cite{MSMiguel} argued that 
fluctuations along the linear boundaries of these strips significantly increase the interfacial free energy due 
to increase in curvature. This mechanism thus lacks instability. According to these authors, instead of a crossover 
from $\alpha=1/3$ to $1$, as in $d=3$, the $2-d$ fluid will encounter a crossover from $1/3$ to $1/2$, the latter 
coming from an interface diffusion mechanism \cite{MSMiguel}. There a crossover from $\alpha=1/3$ to $1$ will 
require large fluctuations, may become possible only at very high temperatures. In the interface diffusion mechanism, 
break-up of strips into droplets is expected \cite{MSMiguel}. In that case, droplet diffusion and collision mechanism of 
Binder and Stauffer (BS) \cite{KBinder3,KBinder4,EDSiggia}, in general valid for off-critical fluid quenches, 
is expected to be operative. This we discuss below. Note that, in case of droplet morphology, the growth 
description cannot be provided via NS equation \cite{AJBray}.
\par
~ As opposed to solid binary mixtures, droplets in fluids are significantly mobile. This gives rise to, in addition to the 
standard evaporation-condensation process, another growth mechanism involving diffusion and collision of droplets. 
These are inelastic collisions, following which the colliding partners merge, forming a bigger droplet and reducing 
the droplet density ($n$). For the decay of $n$, one can write \cite{EDSiggia}
\begin{eqnarray}\label{binder_stauffer}
\frac{dn}{dt}=KD_{\ell} \ell n^{2},
\end{eqnarray}
where $K$ is a constant and $D_{\ell}$ is the diffusion constant for a droplet of size $\ell$. Following 
Stokes-Einstein-Sutherland relation \cite{JPHansen}, $D_{\ell}\ell$ can be treated as a constant. Using $n\sim 1/\ell^d$, 
one obtains \cite{EDSiggia}
\begin{eqnarray}\label{bs_dleng}
\frac{d\ell}{dt} \propto \frac{1}{\ell^{d-1}}.
\end{eqnarray}
Solution of Eq.(\ref{bs_dleng}) provides $\alpha=1/d$. Thus, in $d=2$, BS value of $\alpha$ is same as that for the 
interface diffusion mechanism of San Miguel et al. On the other hand, in $d=3$, BS mechanism provides a value of $\alpha$
coinciding with the LS value. It is expected that the amplitudes will be different for different mechanisms that provide  
same exponents. E.g., in $d=3$ LS mechanism should have a smaller amplitude than the BS one. Direct verification of 
the latter is possible via computer simulations \cite{SRoy1}. Tanaka \cite{HTanaka1,HTanaka2,HTanaka3} argued that 
in high droplet density situation, motion of the droplets will not be random due to inter-droplet interaction. 
However, incorporation of such fact merely changes the growth amplitude. Presence of such interaction can be 
verified by calculating the mean-squared-displacements (MSD) \cite{JPHansen} of the droplets.
\par
~Pictures presented above are primarily meant for kinetics of incompressible liquid-liquid phase separation. 
Nevertheless, they may apply for vapor-liquid phase separation to a good extent. This, however, requires verification.

\section{Models and Methods}
~Kinetics of phase separation in solid binary mixtures at atomistic level have been traditionally studied via 
Monte Carlo (MC) simulations \cite{DPLandau} of the nearest neighbor Ising model
\begin{eqnarray}\label{Ising_hamil}
 H = -J\sum_{<ij>}S_i S_j;~~J>0;~S_{i}=\pm1,
\end{eqnarray}
on regular lattice. Here, an up spin ($S_i=+1$) corresponds to an $A$ particle and a down spin ($S_i=-1$) to a 
$B$ particle. In this model, the kinetics is introduced via the Kawasaki exchange \cite{DPLandau} trial moves 
in which positions of two nearest neighbor particles are interchanged. These moves are accepted according 
to the standard Metropolis algorithm \cite{DPLandau}. At the coarse-grained level, one solves the 
Cahn-Hilliard (CH) equation \cite{AJBray}
\begin{eqnarray}\label{ch_eqn}
\frac{\partial \psi(\vec{r}, t)}{\partial t}=-\nabla^{2}\Big[\psi(\vec{r},t)+\nabla^{2}\psi(\vec{r},t)-\psi^{3}(\vec{r},t)\Big],
\end{eqnarray}
where the order-parameter $\psi$ can be thought of being obtained from the coarse-graining of the Ising spins over a length 
scale of the order of the equilibrium correlation length $\xi$. Thus, the CH equation provides an advantage of accessing 
large effective length scale within practically accessible computation time. 
Eq.(\ref{ch_eqn}) can one way be obtained from the Ising model via a master equation approach with Kawasaki 
kinetics as an input \cite{ISchmidt,SKDas3}. Despite the latter being a direct and rigorous approach, 
in this article we will present a phenomenological derivation. Eq.(\ref{ch_eqn}) is solved on regular lattice, 
typically via Eular discretization technique \cite{SKDas3}. This model, according to the nomenclature of 
Hohenberg and Halperin \cite{MPAllen}, is referred to as the model B.
\par
~In fluids, at coarse-grained level the kinetics is typically studied via model H \cite{MPAllen} which is a combination 
of the CH and the NS equations, the latter taking care of the fluid flow. To describe this model, 
derivation of the CH equation \cite{AJBray} will be of help. Recalling that we are dealing with conserved 
order-parameter, one writes the continuity equation 
\begin{eqnarray}\label{conti_eqn}
 \frac{\partial \psi}{\partial t}= -\vec{\nabla}\cdot\vec{J_c},
\end{eqnarray}
where the current $\vec{J_c}$ comes from the gradient of the chemical potential $-\vec{\nabla}\mu$. The chemical potential 
can be obtained from the functional derivative as
\begin{eqnarray}\label{chem_pten}
\mu=\frac{\delta F[\psi(\vec{r},t)]}{\delta \psi(\vec{r},t)},
\end{eqnarray}
the free energy functional being the Ginzburg-Landau one
\begin{eqnarray}\label{GL_fenrgy}
F=k_B T \int d\vec{r}[-a\psi^2+b\psi^4+c(\vec{\nabla}\psi)^2],
\end{eqnarray}
where the positive coefficients a, b, and c are temperature dependent. Often these coefficients are scaled in such a 
way that the final form is parameter free, as Eq.(\ref{ch_eqn}).
\par
~ The model H equations read \cite{AJBray}, in the incompressible limit ($\vec{\nabla}\cdot \vec{v}=0$), as   
\begin{eqnarray}\label{modelH_eq1}
\frac{\partial \psi}{\partial t}+\vec{v}\cdot\vec{\nabla}\psi=\nabla^2\mu,
\end{eqnarray}
\begin{eqnarray}\label{modelH_eq2}
\rho\frac{D\vec{v}}{Dt}-\nu\nabla^2\vec{v}=-\vec{\nabla}p-\psi \vec{\nabla}\mu.
\end{eqnarray}
In addition to the pure CH and NS equations, there are additional terms in Eqs.(\ref{modelH_eq1}) 
and (\ref{modelH_eq2}), coming from the coupling between the velocity and order-parameter fields. This is justifiable, 
e.g., the current $\vec{J_c}$ in the CH equation is expected to be influenced by the velocity field, becoming 
\begin{eqnarray}\label{current_velo}
 \vec{J_c}=-\vec{\nabla} \mu - \psi\vec{v},
\end{eqnarray}
and the chemical potential gradient should provide a driving force to affect the velocity field in the NS equation. 
Again, these equations can be solved on regular lattice. This model is very much phenomenological and a microscopic 
derivation like the CH equation \cite{ISchmidt} is still lacking. Such an objective can possibly be achieved 
\cite{KKaski} via the construction of free energy functionals by taking density and velocity distributions 
as inputs from molecular dynamics (MD) simulations at various coarse-grained levels, by appropriately 
adjusting the effective inter-particle potential at successive steps. A multi-scale method like this may 
seem obvious but has never been demonstrated. Success of the method may prove useful in justifying 
similar (elegant) phenomenological models in more complex problems like active matter \cite{SRamaswamy}. 
\par
~The incorporation of the coupled terms in Eqs. (\ref{modelH_eq1}) and (\ref{modelH_eq2}), in a sense, 
guarantee the simultaneous growth in the velocity and density fields. While such facts have been 
reported from Lattice Boltzmann simulations \cite{SPuri1,VMKendon}, are lacking in MD simulations even 
though the latter is more direct and accurate. Thus, a microscopic justification of the model, 
particularly understanding of the importance of additional higher order terms, is needed.
\par
~ As can be guessed from the above discussion, at atomistic level, kinetics of fluid phase separation 
is studied via MD simulations \cite{MPAllen,DFrenkel,DCRapaport} in which hydrodynamics can be easily implemented. 
For this purpose, the interatomic interactions are popularly incorporated via the well known Lennard-Jones (LJ) potential
\begin{eqnarray}\label{LJ_pten}
u(r)=4\varepsilon \Big[\Big(\frac{\sigma}{r}\Big)^{12}-\Big(\frac{\sigma}{r}\Big)^6\Big ],
\end{eqnarray}
where $r$ ($=|\vec{r_i}-\vec{r_j}|$) is the distance between two particles at $\vec{r_i}$ and $\vec{r_j}$, $\varepsilon$ 
is the interaction strength and $\sigma$ is the interparticle diameter. In a binary (or multicomponent) mixture, 
$\varepsilon$ and $\sigma$ can be chosen to be different for different combinations of species.
\par 
~In standard MD simulations, equations of motion are solved in discrete time, typically via Verlet velocity algorithm 
\cite{MPAllen,DFrenkel}. Conservation laws of hydrodynamics are perfectly satisfied in microcanonical ensemble using which 
various fluid transport properties are calculated in equilibrium. However, for kinetics of phase separation, 
following a temperature quench, in course of the system's evolution, the potential energy decreases, for energy 
driven phase transitions. In a constant energy ensemble, thus, the kinetic energy, i.e., the temperature increases, 
destroying the objective of the study. Therefore, simulations in canonical ensemble, with appropriate 
hydrodynamics preserving thermostat, become essential. There are a number of good methods for this purpose, e.g., 
dissipative particle dynamics \cite{PNikumen,SDStoyarnov,CPastorino}, multiparticle collision dynamics \cite{AWinkler}, 
Lowe-Andersen thermostat \cite{EAKoopman}, Nos\'{e}-Hoover thermostat \cite{DFrenkel}, etc.
In this article, presented results were obtained via application of the Nos\'{e}-Hoover thermostat (NHT) which will be 
discussed soon.
\section{Simulation Results}
~Binary fluid phase separation has been extensively studied via lattice Boltzmann simulations or simpler solutions of 
the model H equations \cite{VMKendon,SPuri2,CDatt}, for critical quenches. Though MD simulations in this context are 
relatively rare, the theoretical expectations for critical quenches are confirmed to a good degree 
\cite{SKDas1,MLaradji,AKThakre,SAhmad1,SAhmad2,SAhmad3} and reviews are available. Here we focus on vapor-liquid 
phase separation, with particular emphasis on off-critical quenches, providing droplet morphology, 
for which MD simulations are mostly recent \cite{SKDas2,HKabrede,SMajumder3,SRoy1,SRoy2,SRoy3,SWKoch,JMidya}.
\par
~All results were obtained via a modified LJ potential \cite{MPAllen}
\begin{eqnarray}\label{modiLJ_pten}
U(r <r_c)&=&u(r)-u(r_c)-(r-r_c) \frac{du(r)}{dr}\Big|_{r=r_c}, \nonumber 
\\
U(r\geq r_c)&=&0.
\end{eqnarray}
The cut-off radius $r_c$ ($=2.5\sigma$) is introduced for a faster computation. Note that, in critical phenomena 
\cite{MFisher1,HEStanley}, LJ potential being a short-range one, this does not alter the universality class. 
Due to the cut-off and shifting of the potential to $0$ at $r=r_c$, a discontinuity in the force is created 
which may cause non-smooth behavior in energy and thus may be problematic for energy and momentum conservations, 
necessary for preservation of hydrodynamics. This problem is corrected by incorporating the last term in the first part of 
Eq.(\ref{modiLJ_pten}).

\par
~The phase diagram, the primary requirement before studying the kinetics of phase separation, for atomistic models 
can be obtained using both MC and MD simulations. As is well known, equilibrium phase behavior and thermodynamic 
properties are insensitive to ensemble and technicality related to transport mechanism. Thus, MC simulations with smart 
ensembles are advantageous, compared to MD. For vapor-liquid phase transition, typically one uses grandcanonical and 
``Gibbs'' ensemble methods \cite{DPLandau}. With grandcanonical ensemble, other thermodynamic properties, e.g., 
compressibility, can be easily obtained to study its critical singularity.
\par
~As mentioned, an NHT was applied to study dynamics via MD. There one solves the deterministic equations of motion 
\cite{DFrenkel}
\begin{eqnarray}\label{nht_eq1}
m \dot{\vec{r_i}}=\vec{p_i},
\end{eqnarray}
\begin{eqnarray}\label{nht_eq2}
\dot{\vec{p_i}}=-\frac{\delta U}{\delta \vec{r_i}} - \Xi\vec{p_i}, 
\end{eqnarray}
\begin{eqnarray}\label{nht_eq3}
\dot{\Xi}=\frac{\Big[\sum\limits_{i=1}^N (p_{i}^2/m)-3Nk_BT\Big]}{Q}. 
\end{eqnarray}
In Eqs.(\ref{nht_eq1}-\ref{nht_eq3}), $m$ is the particle mass (set equal for all), $\Xi$ is a time dependent drag that 
adjusts its value depending upon the drift of temperature from the assigned value, $N$ is 
the number of particles, $k_B$ is the Boltzmann constant and $Q$ is the coupling strength between the system and 
the thermostat. Essentially, in this scheme one works with a microcanonical ensemble with a modified Hamiltonian. Results 
thus obtained are equivalent to those from the canonical ensemble with the original Hamiltonian \cite{DFrenkel,SDStoyarnov}.
\par
~There exist better hydrodynamics preserving thermostats. But for the present purpose, the NHT appears adequate. The results 
obtained using this thermostat have been tested against other methods. For example, in $d=2$, 
we have compared \cite{JMidya} them with the Lowe-Andersen thermostat (LAT). The LAT is an improvement over 
the basic Andersen thermostat (AT). In the AT randomly chosen particles are assigned new velocities 
(with Maxwell distribution) to keep the temperature at a desired value and thus stochastic in nature, 
like MC methods. In the LAT, while assigning new velocities, conservation of local momentum is appropriately 
taken care of. Further, inside the coexistence region, transport properties of droplets 
(from equilibrium configuration), calculated via NHT, are found  to be in good agreement with the calculations 
using microcanonical ensemble \cite{SRoy2,SRoy3}. Also, in a binary fluid, critical behavior of 
shear viscosity was nicely reproduced via NHT \cite{SRoy4}. In addition, from experience we feel 
that the NHT provides a superior control over temperature, compared to few other better hydrodynamics 
preserving thermostats, particularly in out-of-equilibrium situation. 
\par
~ We will present results \cite{SKDas2,HKabrede,SMajumder3,SRoy1,SRoy2,SRoy3,SWKoch,JMidya} in both $d=2$ and $3$, 
from MD simulations in periodic square or cubic boxes of linear dimension $L$ (in units of $\sigma$). 
All results were obtained after averaging over adequately large numbers of independent initial configurations. 
In the solutions of the equations of motion we have used time discretization $\Delta t=0.005 t_0$, 
$t_0$[$=(m\sigma^2/\varepsilon)^{1/2}$] being an LJ time unit. Note that the critical temperature and the critical number 
density for this model in $d=3$ respectively are $\simeq 0.9\varepsilon/k_B$ and $\simeq 0.3$. All results in $d=3$ 
correspond to quenches from $T=\infty$ to $0.6\epsilon/k_B$. As expected, the value of $T_c$ in $d=2$ is lower 
($\simeq0.5\varepsilon/k_B$). The quench temperature there is $0.25\varepsilon/k_B$.  

\begin{figure}[htb]
\centering
\includegraphics*[width=0.45\textwidth]{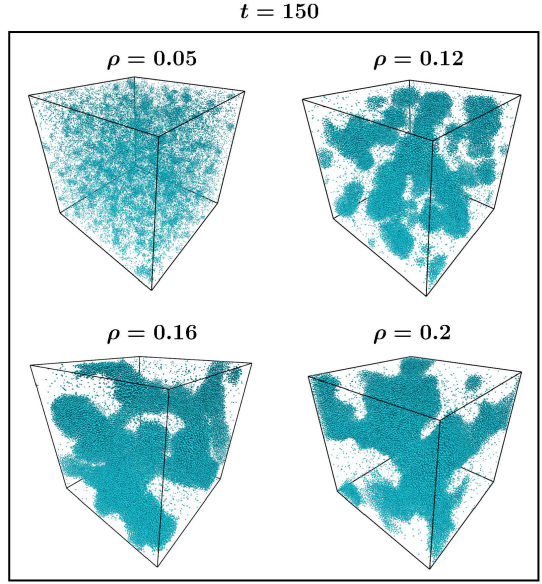}
\caption{\label{fig2} Evolution snapshots for various overall densities $\rho$, obtained from molecular dynamics simulations
of the single component Lennard-Jones model in $d=3$. In all the cases the time, temperature and the system size were
fixed to $t=150$, $T=0.6$ and $L=80$. From J. Chem. Phys. \textbf{139}, 044911 (2013).
}
\end{figure}

\begin{figure}[htb]
\centering
\includegraphics*[width=0.45\textwidth]{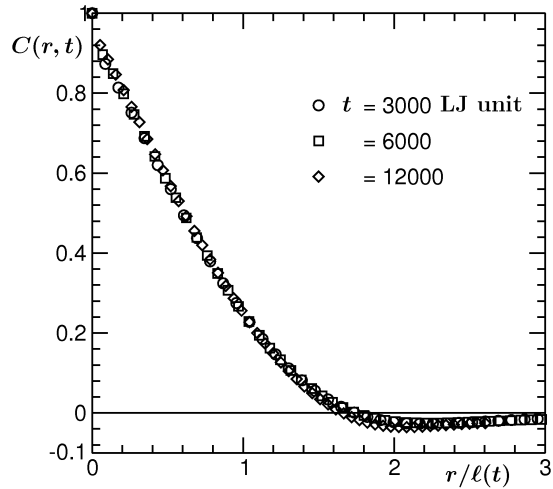}
\caption{\label{fig3} Scaling plot of $C(r,t)$ as a function of $r/\ell(t)$, for $\rho=0.05$. Data from few different times
are included. The results correspond to droplet morphology in $d=3$. From Soft Matter \textbf{9}, 4178 (2013).
}
\end{figure}

\begin{figure}[htb]
\centering
\includegraphics*[width=0.45\textwidth]{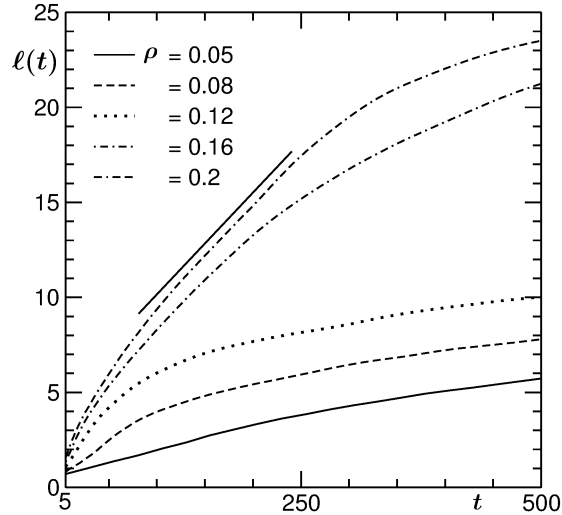}
\caption{\label{fig4} Plots of $\ell(t)$ versus $t$, for different values of $\rho$, in $d=3$. The continuous
straight line corresponds to viscous hydrodynamics growth. From J. Chem. Phys. \textbf{139}, 044911 (2013).
}
\end{figure}

\begin{figure}[htb]
\centering
\includegraphics*[width=0.45\textwidth]{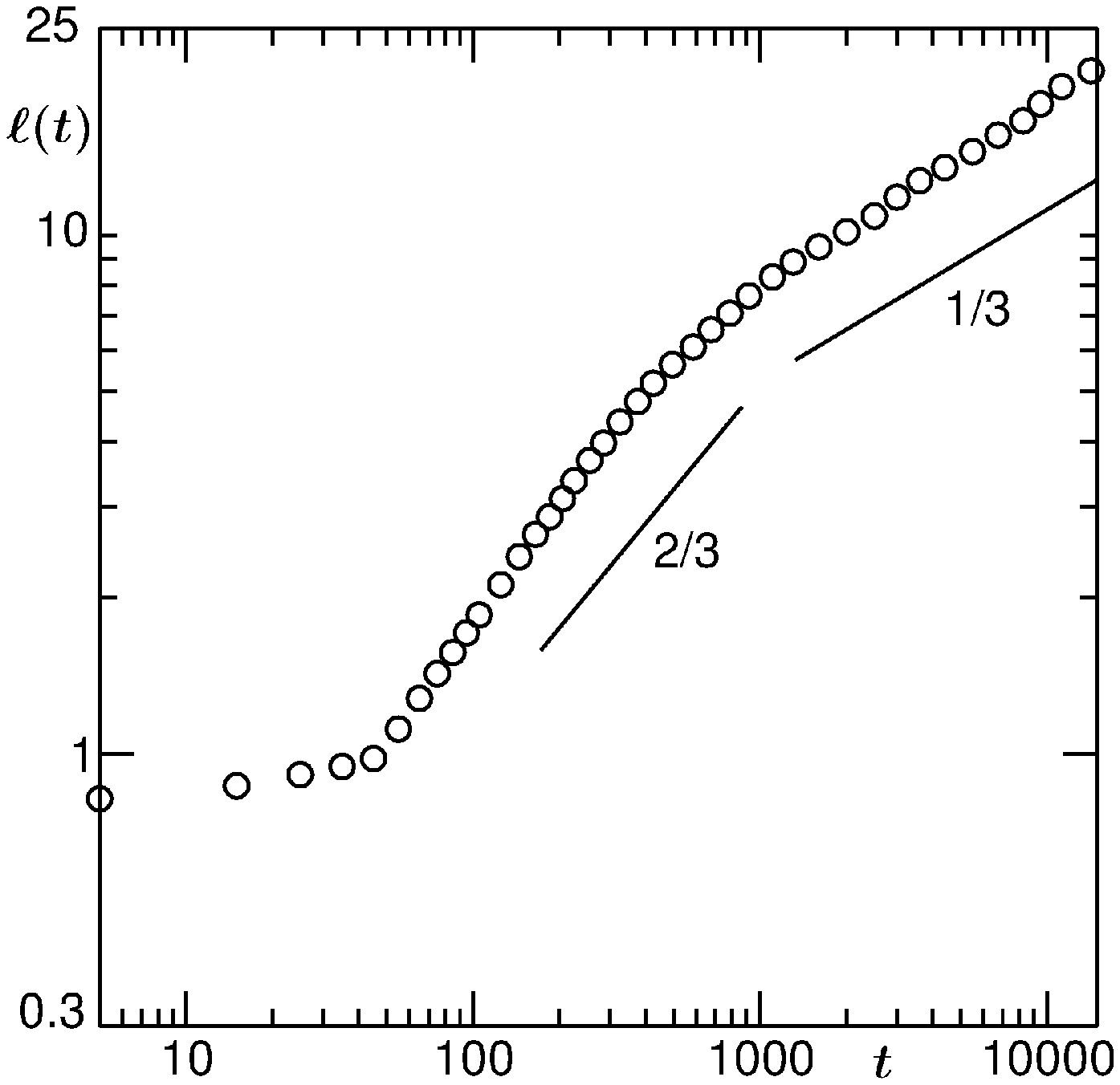}
\caption{\label{fig5} Log-log plot of $\ell$ versus $t$, for $\rho=0.05$, that corresponds to droplet
morphology in $d=3$. The solid lines correspond to power-laws, exponents being mentioned on the figure.
From Phys. Rev. E \textbf{85}, 050602(R) (2012).
}
\end{figure}

\par
~ In Fig.\ref{fig2} we present snapshots from $d=3$ (unless otherwise mentioned, results are from this dimension)
for different overall densities . It is clearly seen that as one approaches the vapor branch of the coexistence curve, 
the morphology becomes more droplet like. Given that we have access to both percolating and droplet structures, a wide 
variety of mechanisms can be checked. 
\par
~In Fig.\ref{fig3} we show a scaling plot of the correlation functions for droplet morphology, using data from different 
times. For this calculation, the order parameter $\psi$ was given a value $+1$ if the density at a point is higher than the 
overall density $\rho$ and $-1$ otherwise. Nice collapse of data, versus $r/\ell(t)$, confirms self-similarity of structure. 
Note here that we have obtained $\ell(t)$ from different methods, from the lengths at which $C(r,t)$ decays to a 
particular value, from the first moment of $P(\ell_d,t)$, as well as by identifying domain boundaries and directly 
counting the number of particles inside them. For the last method, cube (square in $d=2$) root of the average number of 
particles inside the liquid regions will provide $\ell(t)$. For the sake of brevity we present results only from the 
correlation functions.
\par
~In Fig.\ref{fig4} we show plots of $\ell(t)$ vs $t$, for various values of $\rho$. It appears that closer to the critical
density there is a linear growth, after a brief slow regime, before hitting the finite-size effects. The slower 
part at the very beginning can be attributed to the LS behavior \cite{IMLifshitz} and the linear one to the viscous 
hydrodynamic one \cite{EDSiggia,HFurukawa1,HFurukawa2}. Due to the demanding nature of MD simulations, to the best of 
our knowledge, a crossover from viscous to inertial hydrodynamic regime is not yet appropriately observed, using 
this method. However, for intermediate densities, we observed a $t^{2/3}$ behavior even before a linear growth appears. 
This has been checked \cite{SRoy3} via appropriate finite-size scaling analysis \cite{MFisher2} which we do not 
present here. A possible reason for observing the exponent $2/3$ during early period of the hydrodynamic growth 
can be the following. In the inertial regime, due to the large mass of domains, there may be competition between 
growth and break-up. In intermediate density regime, where interconnectedness of the domain morphology is not 
very robust, the break-up becomes easier. For densities very close to the coexistence vapor density, certainly 
the growth is much slower. This we discuss in details below \cite{SRoy1,SRoy2,SRoy3}. 
\par
~In Fig.\ref{fig5} we take a look at the results for $\rho=0.05$ on a double-log scale. The flat behavior of the data 
at the beginning is due to delayed nucleation. A very fast rise after this signals the onset of nucleation, 
following which the data are consistent with a $t^{1/3}$ growth. As mentioned already, in $d=3$ this behavior 
can be due to LS \cite{IMLifshitz,DAHuse1} as well as BS \cite{KBinder3,KBinder4,EDSiggia} mechanism. To select from 
the multiple possibilities we do the following exercise.

\begin{figure}[htb]
\centering
\includegraphics*[width=0.45\textwidth]{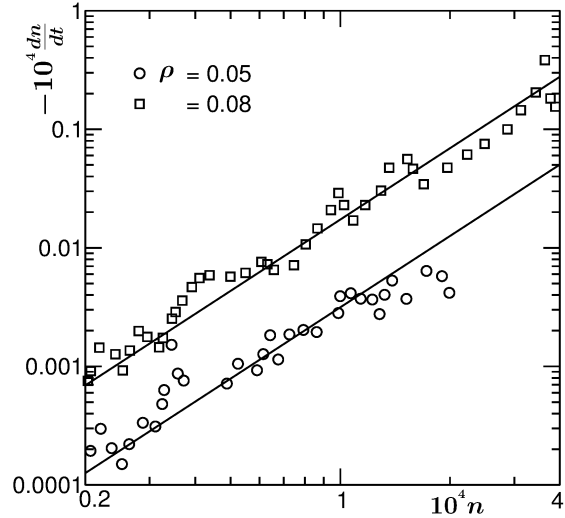}
\caption{\label{fig6} Plots of $dn/dt$ versus $n$, n being the droplet density. Results from two values
of $\rho$, in $d=3$, are included. The solid lines are proportional to $n^2$. For $\rho=0.08$ a multiplicative
factor $5$ was used to separate the data sets from each other. From J. Chem. Phys. \textbf{139}, 044911 (2013).
}
\end{figure}

\begin{figure}[htb]
\centering
\includegraphics*[width=0.45\textwidth]{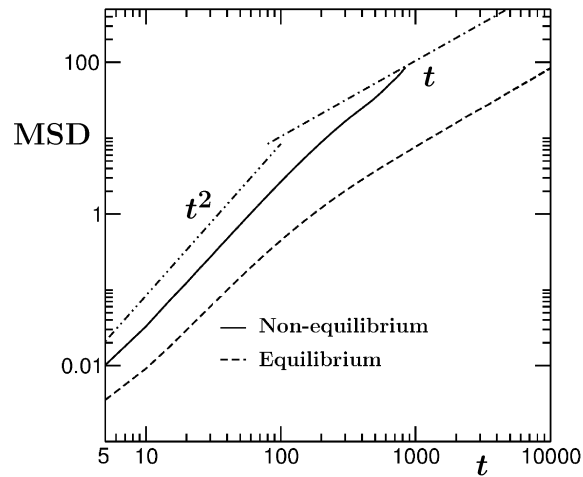}
\caption{\label{fig7} Mean squared displacements for droplets in equilibrium and nonequilibrium situations are plotted
versus time. The droplets are of similar size. The quadratic and linear behavior, corresponding to ballistic
and diffusive regimes, are appropriately marked by different lines. The results corresponds to $d=3$.
From Soft Matter \textbf{9}, 4178 (2013).
}
\end{figure}

\begin{figure}[htb]
\centering
\includegraphics*[width=0.45\textwidth]{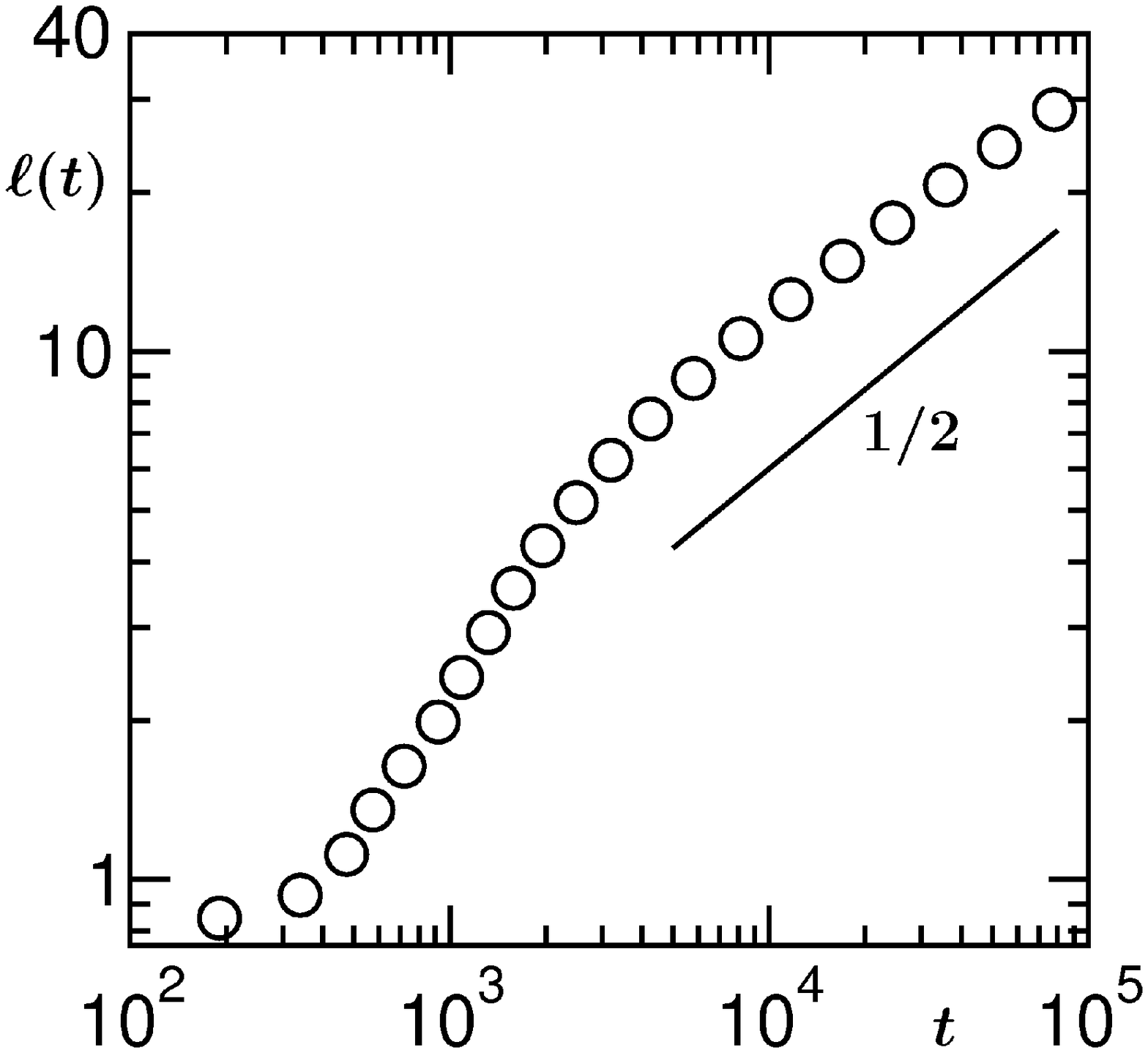}
\caption{\label{fig8} Plot of $\ell$ vs $t$, obtained from the simulations of LJ fluid in $d=2$. The solid line has
$t^{1/2}$ behavior. The results correspond to droplet morphology for overall density $\rho=0.02$.
}
\end{figure}

\par
~ In Fig.\ref{fig6} we show plots of $dn/dt$ versus $n$, for two different low values of $\rho$, 
both providing droplet structure, on a double-log scale. Technical details on the counting of droplets by 
identifying them can be found elsewhere \cite{SRoy1,SRoy2,SRoy3}. Linear look in both the cases confirm 
power-law behavior. The solid lines there are proportional to $n^2$ with which the simulation data are consistent. 
This verifies the starting equation [see Eq.(\ref{binder_stauffer})] for deriving the BS 
\cite{KBinder3,KBinder4,EDSiggia} growth law. Thus, the possibility of a LS \cite{IMLifshitz} mechanism is ruled out. 
We have also checked that between two collisions, sizes of droplets remain same, within minor fluctuations. 
This further discards the importance of LS mechanism in the present context.
\par
~Next, we come to the point of inter-droplet interaction. To understand it we have calculated the MSD \cite{JPHansen} 
of individual droplets \cite{SRoy1,SRoy2} as 
\begin{eqnarray}\label{msd_deff}
{\rm{MSD}}=<(\vec{R}_{CM}(t)-\vec{R}_{CM}(0))^2>,
\end{eqnarray}
where $\vec{R}_{\rm{CM}}(t)$ is the location of the centre of mass of the droplet under consideration at time $t$.
In Fig. \ref{fig7} we show MSD for a typical droplet during nonequilibrium evolution, as a function of time. 
Note that the duration over which such data can be presented is dictated by the average collision interval. This is larger 
for low droplet density and increases with the evolution of the system. In the same graph we showed a corresponding 
plot for a droplet in equilibrium situation, of approximately the same size as the nonequilibrium case. 
The difference between the two cases is significant. On a double-log scale, at late time, the equilibrium 
droplet exhibits linear behavior, confirming Brownian motion. The super-linear behavior in the nonequilibrium 
case is due to inter-droplet interaction, as stated by Tanaka \cite{HTanaka1,HTanaka2,HTanaka3}.
\par
~Finally, in Fig.\ref{fig8} we show $\ell$ versus $t$ plot in $d=2$ for off-critical composition \cite{JMidya}. 
Here clearly an exponent $\alpha=1/2$ is visible. From the calculation of $dn/dt$, in this dimension also we 
have confirmed that the mechanism is BS. For the sake of brevity we do not present these here. For higher 
overall density, the interface diffusion mechanism in this dimension was previously observed \cite{SWKoch}. 
\section{Conclusion}
 ~We have provided a brief review on kinetics of fluid phase separation. A general discussion on various theoretical 
pictures, based on the system dimensionality and domain morphology, is provided. Methodologies related to coarse-grained 
and atomistic models are discussed. Particular emphasis was on the molecular dynamics simulation methods.
\par
~Molecular dynamics results for vapor-liquid transitions are presented 
\cite{SKDas2,HKabrede,SMajumder3,SRoy1,SRoy2,SRoy3,JMidya} in $d=2$ and $3$. These include percolating as well 
as droplet structures. The domain growths from these simulations are observed to be consistent with theoretical 
predictions, despite the fact that most of these predictions are related to phase separation in binary fluids. 
For brevity, we have avoided results on binary fluid mixtures.
\par
~Recently, important results have been obtained with respect to aging in fluid phase separation, for binary fluids 
as well as vapor-liquid systems, via molecular dynamics simulations \cite{SAhmad4,SMajumder4}. These results are 
understood via simple scaling arguments. In addition, effects of disorder in kinetics of fluid phase separation 
have been looked at \cite{SAhmad3} and compared with disordered Ising systems. Lack of space, however, 
restricts us from including these results. 
\par
~Further, there have been significant activities in the area of kinetics in confined geometry 
\cite{RALJones2,SPuri3,HTanaka4,SBastea,SKDas4,SKDas5,MJAHore,KBinder5,PKJaiswal1,PKJaiswal2,EAGJamie}. 
In computational front, good number of reports exist in fluid phase separation 
\cite{SKDas4,SKDas5,MJAHore,KBinder5,PKJaiswal2,EAGJamie} as well. 
However, the understanding of these results are not as complete as pure two- and three- dimensional systems.
\par
~Even though there exists significant agreement between theories and simulations of simple models, the situation is not as 
satisfactory as far as experiments are concerned. Discrepancy between theoretical predictions and experimental observations 
can be attributed to the presence of impurities in real systems as well as to over-simplified pictures in theoretical 
calculations. On the other hand, it is possible to do simulation study of more realistic models \cite{SAhmad3,SJMitchell} 
to obtain better agreement with experiments, for both diffusive and hydrodynamic coarsening, particularly when improved 
computational resources are available.
\vskip 0.5cm
\textbf{Acknowledgement}: The work was partially funded by Department of Science and Technology, Government of India. 
\vskip 0.2cm
${*}$ das@jncasr.ac.in

\end{document}